\documentclass[prl,twocolumn,showpacs]{revtex4}
\usepackage{amsmath}
\usepackage{amssymb}
\usepackage{graphicx}
\usepackage{color}

\begin{document}

\title{Coherent Perfect Absorbers: Time-reversed Lasers}

\author{Y.~D.~Chong}
\email{yidong.chong@yale.edu}

\author{Li Ge}

\author{Hui Cao}

\author{A.~D.~Stone}

\affiliation{Department of Applied Physics, Yale University, New
  Haven, Connecticut 06520}


\begin{abstract}
  We show that an arbitrary body or aggregate can be made perfectly absorbing at discrete frequencies if a precise amount of dissipation is added under specific conditions of coherent monochromatic illumination.  This effect arises from the interaction of optical absorption and wave interference, and corresponds to moving a zero of the elastic S-matrix onto the real wavevector axis. It is thus the time-reversed process of lasing at threshold. The effect is demonstrated in a simple Si slab geometry illuminated in the $500-900 {\rm nm}$ range.  Coherent perfect absorbers are novel linear optical elements, absorptive interferometers, which may be useful for controlled optical energy transfer.
\end{abstract}

\maketitle

A laser is a physical system which, when subjected to an energy flux (pump), self-organizes at a threshold value of the pump to produce narrow-band coherent electromagnetic radiation. In the absence of inhomogeneous broadening and quantum fluctuations, this radiation has zero linewidth.  Above the first lasing threshold, lasers are non-linear systems, but at the first threshold they satisfy a linear wave equation with a negative (amplifying) imaginary part of the refractive index, generated by the population inversion due to the pump \cite{Haken}. In conventional lasers, the gain medium is confined in relatively high quality-factor (high-Q) resonators, and the lasing modes are closely related to passive-cavity modes.  However, recent demonstrations of random lasers \cite{cao_review,TSG0} have shown that the lasing threshold can be reached and coherent lasing obtained in resonators with no high-Q passive-cavity modes. It can be rigorously shown within semiclassical laser theory that the first lasing mode in any cavity is an eigenvector of the electromagnetic scattering matrix (S-matrix) with infinite eigenvalue; i.e.~lasing occurs when a pole of the S-matrix is pulled ``up'' to the real axis by including gain as a negative imaginary part of the refractive index \cite{Beenakker}.  This viewpoint suggests the possibility of the time-reversed process of lasing at threshold \cite{Siegman,Woerdman}.  A specific degree of dissipation is added to the medium, corresponding to a positive imaginary refractive index equal in absolute value to that at the lasing threshold.  The system is illuminated coherently and monochromatically by the time-reverse of the output of a lasing mode, and the incident radiation is perfectly absorbed.  We refer to such an optical system as a coherent perfect absorber (CPA).

Coherent perfect absorption is a general and robust phenomenon related to the analytic properties of the S-matrix. For simplicity, we consider scattering in one or two dimensions, for which the electric field (in the TM polarization) is a scalar obeying the Helmholtz equation:
\begin{equation}
  \left[\nabla^2 + n^2(\vec{r})\, k^2\right] \phi(\vec{r}) = 0.
\end{equation}
Here $k = \omega/c$ (a scalar), $\omega$ is the frequency, and $n = n' + in''$ is the complex refractive index, with $n'' < 0$ for gain and $n'' > 0$ for absorption.  There is an ``external region'', extending from some radius $r_s$ to infinity, where $n = 1$ (the following discussion can be extended straightforwardly to the case of an external dielectric $n_0$).  The field in the external region is a combination of incoming waves $\psi_m^{\textrm{in}}$ and outgoing waves $\psi_m^{\textrm{out}}$:
\begin{equation}
  \phi(\vec{r}) = \sum_m \left[\alpha_m\, \psi_m^{\textrm{in}}(\vec{r}) +
    \beta_m\, \psi_m^{\textrm{out}}(\vec{r})\right]\;, \quad r > r_s.
\end{equation}
The scattering channel amplitudes are related by
$\sum_{m'}S_{mm'}(k) \,\alpha_{m'} = \beta_m$.  With $n''=0$ (lossless medium), $S(k)$ is unitary for real $k$.  Continuing $k$ into the complex plane, $S(k)$ possesses a countably infinite set of poles and zeros, symmetrically placed at $\{k_m^\mp = q_m \mp i \gamma_m  \}$, where $\gamma_m>0$.

When we add gain or dissipation, the zeros and poles of $S(k)$ flow in the complex $k$ plane, but they cannot simply appear or disappear, because $\textrm{arg}\,[\textrm{det}(S)]$ always winds by $2\pi$ around each
zero and $-2\pi$ around each pole.  (We ignore pair creation and annihilation of zeros and poles, which normally does not occur.)   Typically, adding dissipation moves the zeros down, and when there is sufficient dissipation a zero can cross the real axis at some specific value of the wavevector, $\tilde{k}_m$.  We parameterize the dissipation by the imaginary part of the refractive index, $\textrm{Im}\{n_m\} = n''_m > 0 $, and refer to the discrete pairs $\{( \tilde{k}_m, n''_m) \}$ as the CPA zeros.  Radiation incident at each $\tilde{k}_m$ can be completely absorbed {\it if} it corresponds to the specific eigenvector of the S-matrix having eigenvalue zero.  For homogeneous bodies such purely-incoming, real-$k$ radiation modes are the complex conjugate of the constant-flux states first proposed by Kapur and Peierls \cite{KP}, and recently introduced into semiclassical laser theory by T\"ureci \textit{et.~al.} \cite{TSG1}.  We wish to emphasize that the CPA zeros are distinct from absorption resonances of the atomic or molecular medium, which do not require specific illumination conditions.  The CPA process arises from the interplay of interference and absorption: in the presence of specific amounts of dissipation, there exist interference patterns that trap the incident radiation for an infinite time.  Even small rates of single-pass absorption can lead to perfect absorption, and hence media that normally do not absorb radiation well at certain frequencies can be made to do so, albeit within narrow frequency bands.  We will demonstrate this in more detail below, with a simple realizable CPA in Si (silicon). Finally, just as absorption can be enhanced by interference, we show that it can also be reduced by illuminating the CPA with eigenvectors of the S-matrix with constructive interference for escape.

Before moving to specific examples, we discuss a general framework for finding CPA zeros based on the ``R-matrix'' theory of Wigner and Eisenbud  \cite{wigner}.  In this approach, the S-matrix is given by
\begin{equation}
  S(k) = - e^{2ikr_0}\, \left[I - ikR(k)\right]^{-1} \, \left[I +
    ikR(k)\right],
  \label{S matrix}
\end{equation}
where $r_0 > r_s$ is an arbitrarily-chosen boundary far from the origin, and $R(k)$ (the R-matrix) takes the form
\begin{eqnarray}
  R_{mm'}(k) &=& \sum_{\alpha\alpha'=1}^\infty \phi_\alpha^m
  \,F_{\alpha\alpha'}^{-1}\, \phi_{\alpha'}^{m'} \\
  F_{\alpha\alpha'} &=& (k_\alpha^2-k^2)\,\delta_{\alpha\alpha'}
  - ik^2 \gamma_{\alpha\alpha'},
\end{eqnarray}
where $\phi_{\alpha}^m \in \mathbb{R}$ is the $\alpha$-th Wigner-Eisenbud
basis function evaluated at $r=r_0$ and decomposed into the $m$-th channel \cite{unitary}, while $k_\alpha \in \mathbb{R}$ is the eigenvalue of that basis function.  The dissipation matrix, $\gamma_{\alpha\alpha'} = \int d^dr\, \textrm{Im}\,(n^2)(\vec{r})\, \phi_\alpha(\vec{r})\, \phi_{\alpha'}(\vec{r})$, is real and positive-definite.

\begin{figure}
\centering
\includegraphics[width=0.37\textwidth]{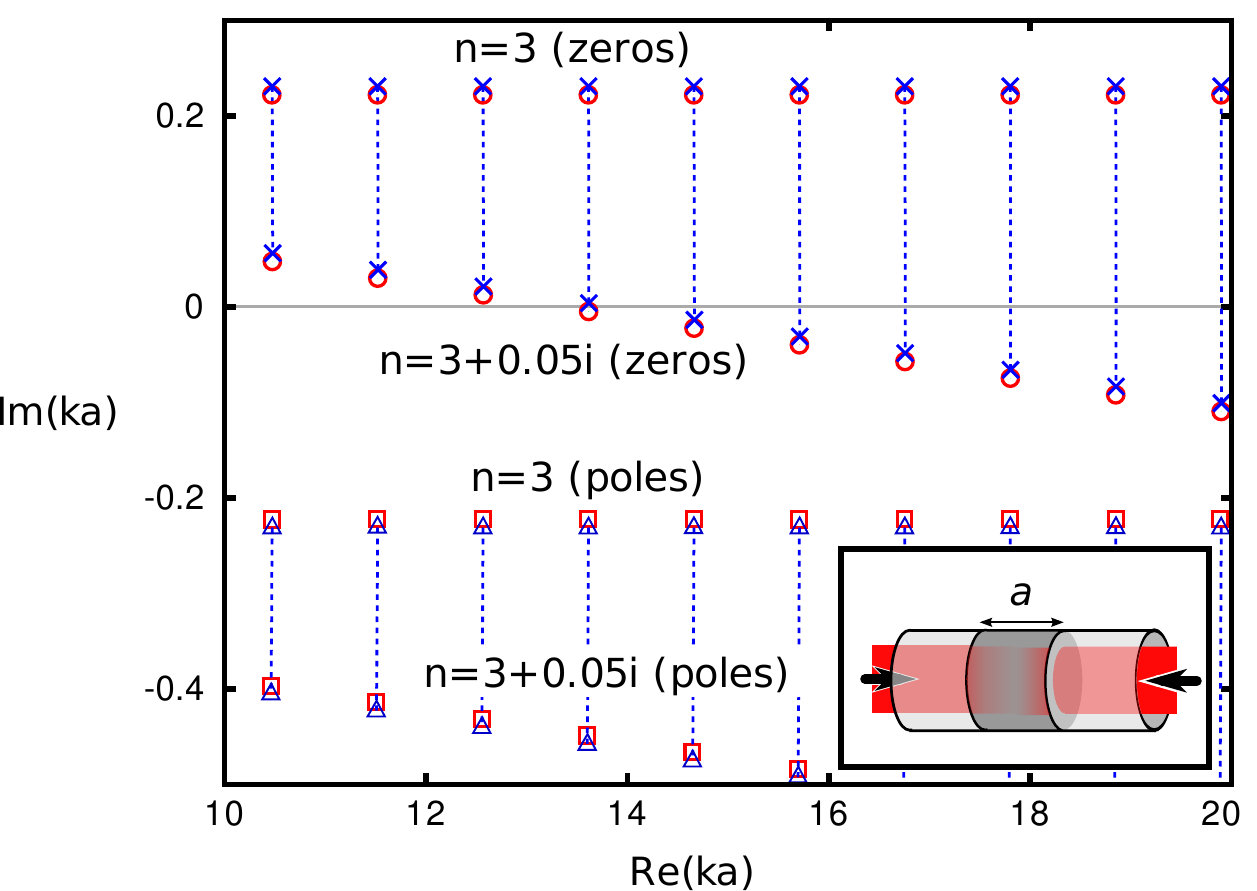}
\caption{(color online) Motion of exact S-matrix zeros (blue crosses) and poles (blue triangles) in the complex-$k$ plane, as dissipation increases from zero, for a one-dimensional slab of length $a$ and uniform index $n$.  Also shown are the zeros (red circles) and poles (red squares) from Equation~(\ref{polemot}), the single R-matrix pole approximation.  Inset: schematic of system.}
\label{polemotion}
\end{figure}

When $Q\gg1$, each S-matrix zero/pole is determined by approximating the R-matrix by a single term $\alpha$; in this case the zeros and poles of the S-matrix occur at
\begin{equation}
  \left(1 + i \gamma_{\alpha\alpha}\right) k^2\, \mp i \varphi_\alpha \,k\, -
  k_\alpha^2 = 0,
  \label{polemot}
\end{equation}
where $\varphi_\alpha = \sum_m (\phi_\alpha^m)^2 >0$.  Without dissipation ($\gamma_{\alpha\alpha} = 0$), this implies that all zeros have positive imaginary parts, as already noted.  Furthermore, zeros cross the real axis exactly when $\gamma_{\alpha \alpha} = \varphi_\alpha /k_\alpha$, at frequency $k = k_\alpha$.  In Fig.~\ref{polemotion}, we show the exact pole motion, for the case of the two-port CPA (see inset and discussion below), finding excellent agreement with the single-pole R-matrix prediction, even though the cavity only has $Q\sim30$.

\begin{figure}
\centering
\includegraphics[width=0.378\textwidth]{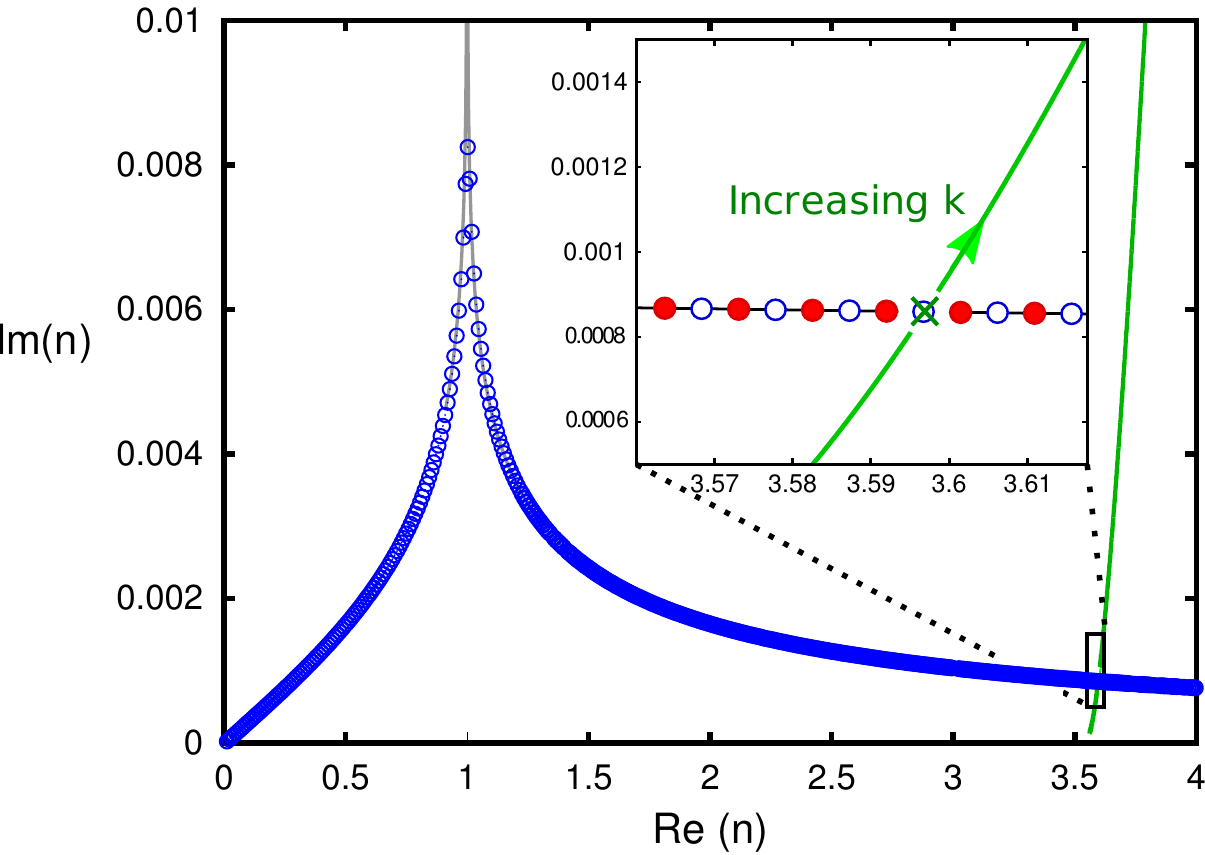}
\caption{(color online) Complex refractive indices for the uniform dielectric slab as a coherent perfect absorber, for $ka = 664.7$.  Parity-even solutions are shown as hollow blue circles; parity-odd solutions are omitted for clarity.  The green curve shows the refractive index of Si at different frequencies.  Inset: parity-even solutions (hollow blue circles) and parity-odd solutions (filled red circles) in the region $3.55 < \textrm{Re}(n) < 3.62$; the green cross shows the index of Si at $ka = 664.7$, $a = 100\, \mu\textrm{m}$ (i.e.~$\lambda=945.3\, \textrm{nm}$). }
\label{slabzeros}
\end{figure}

The simplest possible CPA is a single port reflector, similar to the
``critically-coupled fiber-resonator'' systems widely studied in integrated optics \cite{ccr1,ccr2,caveat}.  However, important properties of the CPA are only revealed when there are multiple input ports and hence non-trivial eigenvectors.  We therefore study a two-port case to illustrate the concept fully.  Consider a finite one-dimensional slab of thickness $a$ and uniform refractive index $n$, with two input channels for each $k$ corresponding to incident radiation from left and right.  It is straightforward to calculate the $2\times2$ S-matrix for arbitrary complex index $n$, and to show that it has a zero eigenvalue when
\begin{equation}
  e^{inka} = \pm \frac{n-1}{n+1}.
  \label{nvals}
\end{equation}
When $ka \gg 1$, we can find an infinite number of discrete solutions of this equation, $n_\nu
= n'_\nu + i n''_\nu$, as
\begin{eqnarray}
  n'_\nu &\approx& \frac{\pi \nu}{ka}\,,\quad\nu = 1,2,3,\cdots \label{n1} \\ n''_\nu
  &\approx& \frac{1}{ka}\, \ln\left(\frac{n'_\nu+1}{n'_\nu-1}\right).\label{n2}
\end{eqnarray}
In Fig.~\ref{slabzeros}, we plot the solutions for $ka = 664.7$. Note that we have here restated the CPA problem so as to find the $\{n_\nu (k) \}$ which produces a zero of the S-matrix for a fixed $k$.  This can be achieved by letting both the real and imaginary parts of $n_\nu$ vary; earlier we varied the imaginary part $n''$ at fixed $n'$, leading to zeros at different $k$-points.  This is a useful reformulation because it suggests how a CPA can be achieved in practice.  A real material has a frequency-dependent $n(k)$, and when the material is fabricated and illuminated in a 2-port configuration, one may realize a CPA by scanning $k$ and looking for coincidences, i.e. $n(k) \approx n_\nu (k)$ for some integers $\nu$.

\begin{figure}
\centering
\includegraphics[width=0.39\textwidth]{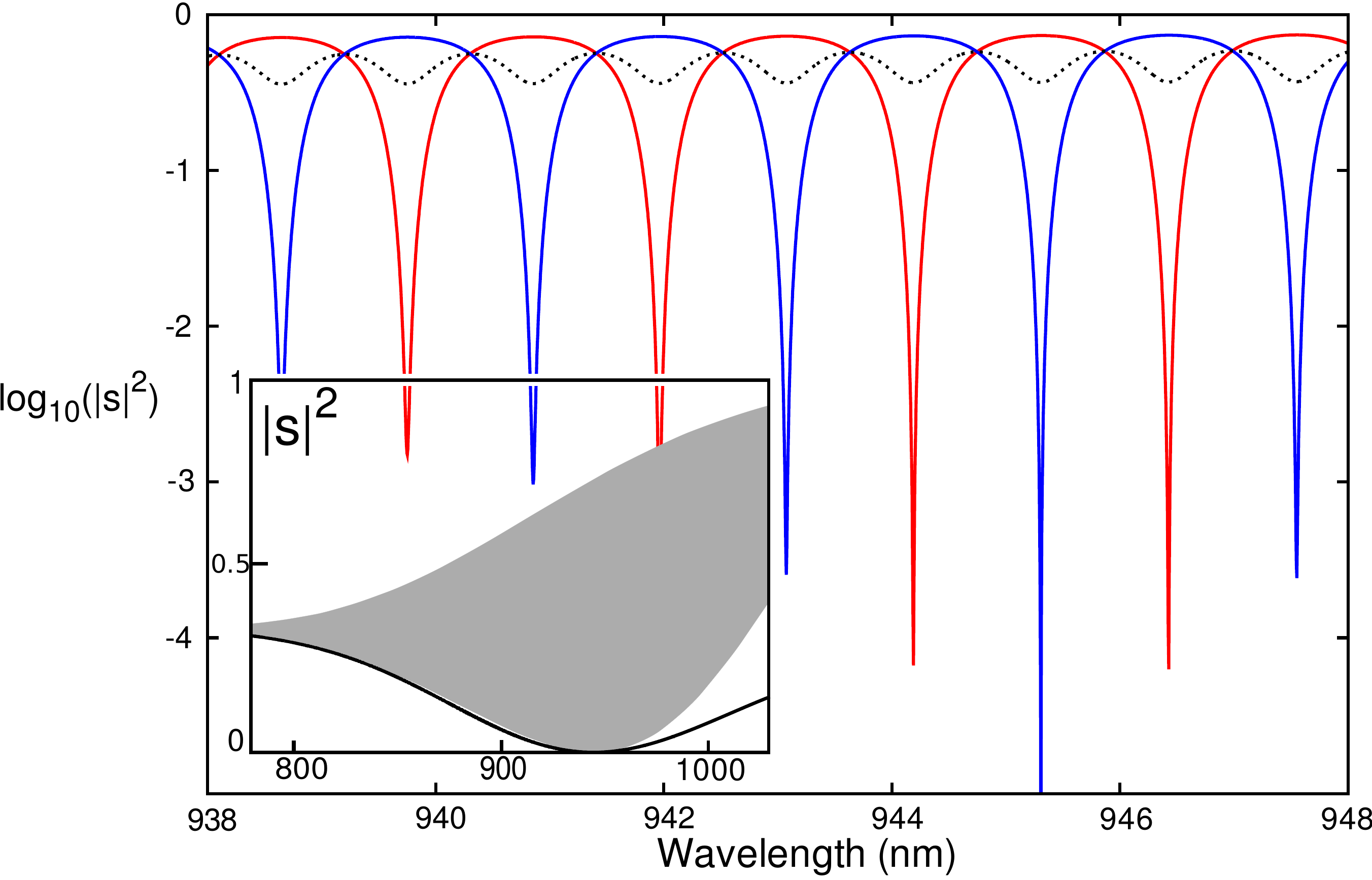}
\caption{(color online) Semi-log plot of normalized output intensities (i.e.~$|s|^2$, where $s$ are S-matrix eigenvalues) vs.~the wavelength $\lambda = 2\pi/k$, for a 100 $\mu$m Si slab.  Solid lines show $\log_{10}(|s|^2)$ for a parity-even (blue) or parity-odd (red) eigenmode.  The dashed line shows $2(|r|^2 + |t|^2)$, the total output intensity when the two input beams are incoherent.  Inset: upper and lower bounds for $|s|^2$ over a wide range of $\alpha$.  The gray area shows the actual bounds, and the solid black line show the approximate lower bound from Equation~(\ref{ibound}). }
\label{siplot}
\end{figure}

In an indirect bandgap semiconductor such as Si, $n''$ increases smoothly from very small values as $k$ increases past the bandgap, while $n'$ changes little \cite{Adachi}.  For frequencies near the bandgap, $\partial n''/ \partial k \gg \partial n''_\nu/ \partial k$ and $\partial n'/ \partial k \ll \partial n'_\nu/ \partial k$; thus it is possible to vary k so as to pass very close to several CPA zeros.  Like a bound state, a CPA zero has no intrinsic width; the scattered intensity near a CPA zero is $I_\textrm{min} \sim c_0 |n(k) - n_\nu(k)|^2$, where  $c_0$ is of order unity (neglecting the spectral width of the incident radiation).  Fig.~\ref{siplot} shows the S-matrix eigenvalue intensities for a slab of undoped Si, with $a$ = 100 $\mu$m.  For this system, $|n(k) - n_\nu(k)|^2 \lesssim 10^{-5}$ for the optimal $(\nu, k)$. As we see, deep CPA minima are achieved in the neighborhood of $\lambda = 945 {\rm nm}$, with a maximum intensity contrast $ \sim 50$ dB, and more than ten substantial minima are visible in the range $ 938 {\rm nm} < \lambda < 954  {\rm nm}$.  Due to the mirror symmetry of the uniform slab, the two eigenmodes are parity eigenstates.  The $\pm$ in Eq.~(\ref{nvals}) determines whether the perfect-absorber condition is satisfied by the symmetric or antisymmetric mode.   The location of these minima is $a$-dependent and hence tunable within a given material; we can derive a tight lower bound for the S-matrix eigenvalue intensities,
\begin{equation}
  |s(a)|^2  \ge \left[2\, \frac{(n'^2-1) \, \sinh(n'' ka) - 2n'}{(n'+1)^2\,e^{n''
        ka} + (n'-1)^2\,e^{-n'' ka}} \right]^2,
  \label{ibound}
\end{equation}
which goes to zero at the $n$ values given in (\ref{nvals}) and locates the interesting operating regions. For Si, the optimal wavelength occurs at around 750 nm for $a$ = 10 $\mu$m, and around 1000 nm for $a$ = 150 $\mu$m; the exact value depends on $n(k)$, which in turn depends on the doping.  If $a \gtrsim 50 \mu$m, the spacing between CPA zeros, given in (\ref{n1}), becomes small, and the material index always passes close to one or more of them without fine-tuning $a$, making the behavior shown in Fig.~\ref{siplot} robust. Simulations of CPAs of this type for other indirect bandgap semiconductors  such as GaP show similarly good results.

Perfect absorption occurs only under the correct coherent two-channel illumination conditions, corresponding to the eigenvector of the S-matrix with eigenvalue zero.  The effect arises from a combination of interference and dissipation: the reflected part of the first incident beam interferes destructively with the transmitted part of the second incident beam, and vice versa, and therefore the radiation is trapped in an interference pattern within the slab and lost entirely to dissipation.  One sees from Fig.~\ref{siplot} that the other, orthogonal eigenmode has maximal $|s|^2$, and in fact the material absorbs significantly less energy than if it is incoherently illuminated from both sides.  In this mode, the reflected part of each beam interferes constructively with the transmitted part of the other beam, causing the radiation to escape the slab more quickly than in the incoherent case.

To illustrate the role of interference in the CPA, we write down
the transfer matrix, a $2\times2$ matrix with unit determinant:
\begin{equation}
  T(n,k) = \frac{1}{t} \begin{bmatrix} t^2-r^2 & r \\ -r & 1
  \end{bmatrix},
  \label{Tmatrix}
\end{equation}
where $r(n,k)$ and $t(n,k)$ are the reflection and transmission amplitudes for a single wave of unit amplitude incident from either direction.  Perfect absorption occurs when $T_{11} = 0$, i.e.~when $r^2 = t^2$.  (Furthermore, we can show that the non-zero eigenvalue, corresponding to the minimally-absorbed mode, is $2t$.)  In order for the two output beams to interfere destructively in the manner described above, not only must they have equal intensities ($|r|^2 = |t|^2$), they must also have the correct relative phase.  This is impossible to satisfy when $n$ is real, as expected from energy conservation.  In the T-matrix analysis, this property is based on the impossibility of establishing the correct interference pattern; when $n$ is real, $T^* = T^{-1}$, and consequently (\ref{Tmatrix}) implies that even if we achieve the condition $|r|^2 = |t|^2$, there is always a $\pm \pi/2$ phase difference between $r$ and $t$, so $r^2 \ne t^2$.

The above analysis applies to any 2-port system with mirror symmetry in the index variation.  It can be extended to arbitrary 2-port systems, for which the general transfer matrix has the form
\begin{equation}
  T = \frac{1}{t} \begin{bmatrix}t^2 - r_\ell r_r & r_r \\ - r_\ell & 1
  \end{bmatrix},
\end{equation}
where $r_\ell$ and $r_r$ are the reflection amplitudes for a wave of unit
amplitude incident from either the left and right respectively, and $t$ is the
transmission amplitude for either direction of incidence.  Then the perfect
absorption condition can be written as $r_\ell r_r = t^2$, and again this
cannot be satisfied if the refractive indices are all real.

One interesting property of the mirror-symmetric system is that we can change the output intensity simply by changing the relative phase of the input beams.  For a phase-modulated input, $[1,\pm e^{i\phi}]$, the output intensities are equal in each channel, with the form
\begin{equation}
  I = I_0 \, \sin^2\left(\frac{\phi}{2}\right).
  \label{I0}
\end{equation}
For the uniform slab, $I_0 = |(n^2-1)/(n^2+1)|^2$, where $n$ is the special refractive index satisfying (\ref{nvals}).  The mirror-symmetric CPA therefore functions as a compact absorbing interferometer.  Unlike an ordinary interferometer, it does not deflect the input beams into another output channel, but causes it to be absorbed entirely within the material.  The CPA may therefore be good for applications in which optical energy needs to be delivered to a device in a precise pulse shape or sequence, which could be controlled by modulating the relative beam phase in Eq.~(\ref{I0}).  Once the CPA has absorbed energy, it must flow out in some manner, e.g. as heat or as fluorescence, the latter being less likely to be useful or desirable.  Hence direct bandgap materials, such as GaAs, which make good lasers, make poor CPAs since much of the absorbed light will be re-radiated.  In contrast, indirect bandgap systems such as Si, which are difficult to make into lasers, retain the energy as heat or as electron-hole pairs, leading to an enhanced photovoltaic response as CPAs. We emphasize however that if the incident radiation is broadband, then the response averages out to approximately the incoherent result, due to the oscillation of the absorption in Fig.~\ref{siplot}.

\begin{figure}
\centering
\includegraphics[width=0.39\textwidth]{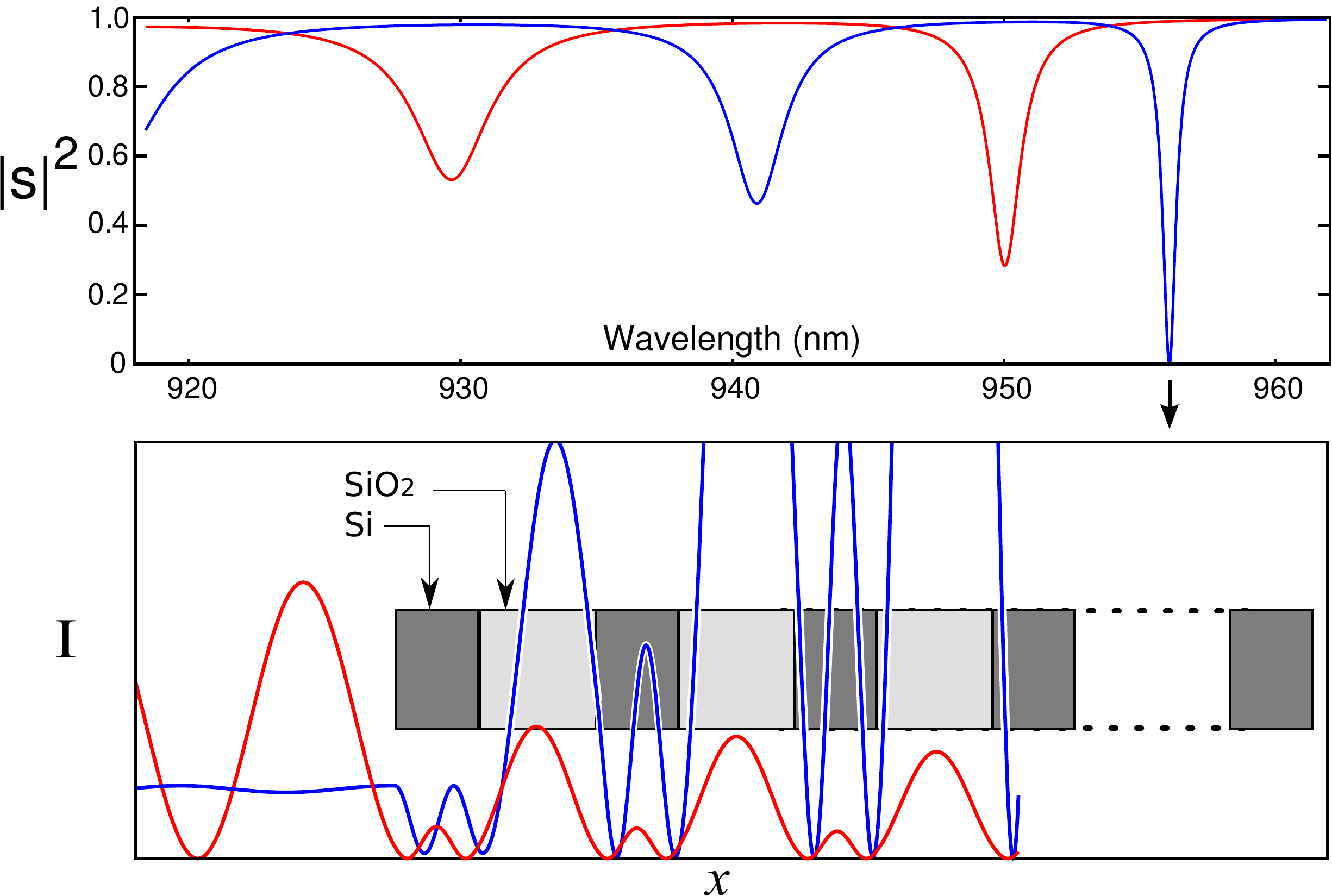}
\caption{(color online) Plot of S-matrix eigenvalue intensities $|s|^2$, as a function of input wavelength, for a
mirror-symmetric  ``Bragg CPA"  consisting of 20 Si slabs of width 188.5 nm, separated by SiO$_2$ slabs of width 266 nm. The refractive index of SiO$_2$ ($n = 1.46$) is real to a good approximation in this wavelength range, so absorption only occurs in the Si.  A large ($\sim$ 99\%) absorption contrast can be observed at 956 nm, close to one of the system's photonic band edges.  Inset: the intensity of the two eigenmodes, as a function of $x$, close to the edge of the system, at 956 nm.  }
\label{dbrplot}
\end{figure}

In applications it will often be useful to have a large contrast between the perfectly absorbed and incoherent or reduced absorption illumination conditions.  According to (\ref{I0}), the intensity contrast between the maximally and minimally absorbing eigenmodes in a uniform slab is limited by $n$.  For the  Si slab of Fig.~\ref{siplot}, the contrast is $\sim 0.75$.  (In particular, $I_0 < 1$ for all $n$, because some field is always present in the absorbing material.)  It is possible to increase the contrast with a non-uniform system, such as a periodic array of slabs (i.e., a distributed Bragg reflector or one-dimensional photonic crystal).  One such system, consisting of Si and non-absorbing SiO$_2$ slabs, is shown in Fig.~\ref{dbrplot}.  It exhibits a much larger absorption contrast than the uniform slab: the S-matrix eigenvalues at the 956 nm absorption resonance have intensities $|s_1|^2 = 5\times10^{-4}$ (symmetric) and $|s_2|^2 = 0.988$ (antisymmetric).  The absorption in the antisymmetric mode is exceptionally low because the field inside the slab is concentrated in the SiO$_2$ regions, which are non-absorbing.

Finally, we note the CPA is probably easiest to realize in a waveguide geometry, where it is relatively easy to control individual mode amplitudes.  In a free-space geometry, it may be challenging to generate the input corresponding to the desired S-matrix eigenvector, due to the proliferation of channels.

This work was partially supported by NSF Grant Nos.~DMR-0808937 and DMR-0908437.  We thank Hakan T\"ureci, Marin Solja\v{c}i\'{c}, John Joannopoulos, Eric Ippen, Qinghai Song, and Heeso Noh for helpful discussions.


\begin{thebibliography}{99}
\bibitem{Haken} H.~Haken, \textit{Laser Light Dynamics} (North Holland, 1986).
\bibitem{cao_review} H.~Cao, J. Phys. A: Math. Gen.~{\bf 38}, 10497 (2005).
\bibitem{TSG0} H.~E.~T\"ureci, L.~Ge, S.~Rotter, and A.~D.~Stone,
  Science {\bf 320}, 643 (2008).
\bibitem{Beenakker} K.~M.~Frahm, H.~Schomerus, M.~Patra, and C.~W.~Beenakker, Europhys.~Lett.~{\bf 49}, 48 (2000).
\bibitem{Siegman} A.~E.~Siegman, Phys.~Rev.~A~{\bf 39}, 1253 (1989).
\bibitem{Woerdman} W.~A.~Hamel and J.~P.~Woerdman, Phys.~Rev.~A~{\bf 40}, 2785 (1989).
\bibitem{KP} P.~L.~Kapur and R.~Peierls, Proc.~Roy.~Soc.~London~{\bf A} 166,
  277 (1938).
\bibitem{TSG1} H.~E.~T\"ureci, A.~D.~Stone and B.~Collier, Phys.~Rev.~A {\bf
  74}, 043822 (2006).
\bibitem{wigner} E.~P.~Wigner and L.~Eisenbud, Phys.~Rev.~{\bf72}, 606
  (1946).
\bibitem{unitary} We choose the basis set corresponding to the dissipationless system, i.e. $n = n'$, so that the imaginary part of the R-matrix is expressed explicitly in terms of ${\rm Im} \{ n^2 \}$.
\bibitem{ccr1} M.~Cai, O.~Painter, and K.~J.~Vahala, Phys.~Rev.~Lett.~{\bf 85} 74 (2000).
\bibitem{ccr2} J.~R.~Tischler, M.~S.~Bradley, and V.~Bulovi\'c, Opt.~Lett. 31, 2045 (2006).
\bibitem{caveat} In a fiber critically coupled to a resonator, part of the loss is usually radiative and not due to absorption.
\bibitem{Adachi} S.~Adachi, \textit{Optical Constants of Crystalline and
  Amorphous Semiconductors}.  Springer, 1999.
\end{thebibliography}
\end{document}